\documentclass[aps,twocolumn,superscriptaddress,floatfix,nofootinbib,prl]{revtex4-1}

\usepackage{lipsum}
\usepackage{graphicx}
\usepackage{amsmath,amssymb}
\usepackage{braket}
\usepackage{mathtools}
\usepackage{hyperref}

\usepackage{color}
\usepackage[normalem]{ulem}
\usepackage{amsfonts}

\usepackage{pdfpages} 
\makeatletter
\AtBeginDocument{\let\LS@rot\@undefined}
\makeatother

\newcommand{\cosq}[1]{\cos\left(2\pi\varphi #1\right)}

\begin{document}
\title{Universality and Quantum Criticality in Quasiperiodic Spin Chains}

\author{Utkarsh Agrawal}
\affiliation{Department of Physics, University of Massachusetts, Amherst, MA 01003, USA}
\author{Sarang Gopalakrishnan}
\affiliation{Department of Physics and Astronomy, CUNY College of Staten Island, Staten Island, NY 10314;  Physics Program and Initiative for the Theoretical Sciences, The Graduate Center, CUNY, New York, NY 10016, USA}
\author{Romain Vasseur}
\affiliation{Department of Physics, University of Massachusetts, Amherst, MA 01003, USA}

\begin{abstract}

Quasiperiodic systems are aperiodic but deterministic, so their critical behavior differs from that of clean systems as well as disordered ones. Quasiperiodic criticality was previously understood only in the special limit where the couplings follow discrete quasiperiodic sequences. Here we consider generic quasiperiodic modulations; we find, remarkably, that for a wide class of spin chains, generic quasiperiodic modulations flow to discrete sequences under a real-space renormalization group transformation. These discrete sequences are therefore fixed points of a \emph{functional} renormalization group. This observation allows for an asymptotically exact treatment of the critical points. We use this approach to analyze the quasiperiodic Heisenberg, Ising, and Potts spin chains, as well as a phenomenological model for the quasiperiodic many-body localization transition.

\end{abstract}
\maketitle
\vspace{1cm}

Quenched randomness has dramatic effects on the thermodynamics and response of one-dimensional quantum systems. Infinitesimal randomness can completely change the critical properties of quantum phase transitions, if the correlation length exponent $\nu$ violates the Harris criterion $\nu>2$~\cite{harris_effect_1974}.
For a wide variety of models of quantum magnetism, the resulting random critical points are of the ``infinite-randomness'' type, for which asymptotically exact renormalization-group methods exist~\cite{fisher_random_1992,PhysRevB.50.3799,PhysRevB.51.6411,IGLOI2005277}. Many properties of these critical points can be understood by central limiting arguments that rely on the uncorrelated nature of the disorder potential. 
Such infinite-randomness fixed points also seem to exist in two dimensions, although the theory is less controlled in this case~\cite{PhysRevB.61.1160}.

Many systems of current interest---such as quasicrystals~\cite{PhysRevLett.53.1951,PhysRevLett.53.2477,PhysRevLett.55.1768}, twisted bilayer graphene~\cite{Bistritzer12233,Cao:2018aa,Cao:2018ab}, and cold atoms in bichromatic laser potentials~\cite{Roati:2008aa,Deissler:2010aa,Schreiber842,PhysRevX.7.011034,PhysRevX.7.041047}---are inhomogeneous, but with quasiperiodic rather than random modulation of the couplings. Quasiperiodic patterns are deterministic and have long-range spatial correlations: thus the central limiting arguments that describe random critical points fail in the quasiperiodic case. 
Clean critical points are more stable to quasiperiodic than to random potentials: the most general stability criterion, due to Luck, is $\nu \geq 1$ (in one dimension)~\cite{luck_classification_1993}. The Heisenberg and Potts models violate even this weaker criterion, so when their couplings are quasiperiodically modulated they flow to a critical point dominated by the modulation. The study of such quasiperiodic quantum critical points is in its infancy; while certain special cases (in which the couplings come from a binary substitution sequence) have been analyzed using renormalization-group (RG) methods~\cite{kohmoto83, satija_quasiperiodic_1988,levitov, igloi_random_1998,vidal_correlated_1999,hermisson_aperiodic_1999,hida_new_2004,vieira_aperiodic_2005}, the case of generic quasiperiodic modulation has only been addressed for free fermions~\cite{wilkinson, PhysRevX.7.031061,crowley_critical_2018,crowley_quasiperiodic_2018}.

The central result of this work is that, for a wide class of models, generic quasiperiodic modulations flow under renormalization to discrete substitution sequences. These sequences are \emph{attractors} in the space of quasiperiodic modulations. Since generic patterns flow to substitution sequences, for which asymptotically exact real-space RG schemes exist, we can construct asymptotically exact descriptions of generic quasiperiodic quantum critical points, and describe many of their properties (such as critical exponents) analytically. 
In what follows, we explain why (and under what conditions) substitution sequences are attractors, using the illustrative example of the Heisenberg spin chain; we then extend our analysis to the Ising and Potts models, and finally to a toy model of the MBL transition~\cite{zhang_many-body_2016,goremykina_analytically_2019}.

\emph{Quasiperiodic Heisenberg chain}.--- Although our results are very general and apply to a variety of one-dimensional systems, for concreteness we will illustrate the approach on a paradigmatic example of quantum magnetism: the antiferromagnetic spin-$\frac{1}{2}$ Heisenberg spin chain
\begin{equation}
H=\sum_i J_i \vec{S}_i \cdot \vec{S}_{i+1}, \label{eqXXX}
\end{equation}
with $J_i>0$. In the clean case $J_i=J$, this spin chain is gapless and is described at low energy by a $SU(2)$ symmetric Luttinger liquid with Luttinger parameter $g=\frac{1}{2}$. Disorder in the $J_i$ couplings is a relevant perturbation~\cite{giamarchi_anderson_1988} that leads to a quantum critical random-singlet state~\cite{fisher_random_1992}. In the random case, the low-energy physics can be captured by a strong disorder real space RG with the following iterative procedure~\cite{ma_random_1979,dasgupta_low-temperature_1980}: one identifies the strongest remaining coupling $J_i$, forms a singlet out of the spins $i, i+1$ (thus eliminating them from the problem), and generates a new effective coupling $J_{\rm eff}=\frac{J_{i-1}J_{i+1}}{2 J_i}$---at second order in perturbation theory---between spins $i - 1$ and $i + 2$. This procedure is accurate so long as $J_i \gg J_{i-1},J_{i+1}$; in the random case, the ratio of neighboring couplings flows to infinity, so the procedure is asymptotically exact~\cite{PhysRevB.50.3799}.

Here, we are interested instead in quasiperiodic modulations of the couplings, with $J_i=f(i)$ with $f(x) = f(x + \varphi^{-1})$ for irrational $\varphi$ [in the bulk of this text we take $\varphi =  \frac{1+\sqrt{5}}{2}$]. We take $f >0$ to be a general smooth function with a smooth logarithm. To understand whether this perturbation is relevant at the Heisenberg ``critical point'' we recall that the Heisenberg chain is a critical point separating two inequivalent dimerized phases, which occur when the even and odd bonds have different strengths. 
The correlation length exponent for the dimerization transition is $\nu = 2/3 < 1$; thus, weak quasiperiodicity is relevant~\cite{vidal_correlated_1999,hida_density_1999,vidal_interacting_2001} by the Harris-Luck criterion~\cite{luck_classification_1993}, and the system flows to a quasiperiodicity-dominated fixed point.

\emph{Flow to discrete sequences}.---We now apply the standard real-space decimation procedure (described above for the random case) to this quasiperiodic model. It is convenient to write the decimation rule as
\begin{equation}
\ell'_i=\ell_{i-1}-\ell_i+\ell_{i+1}+c, \label{eqRGrule}
\end{equation} 
 where we have defined $\ell_j=-\ln{J_j}$, $c=\ln 2$ (for the Heisenberg model), and $\ell_i = \min \lbrace \ell_j\rbrace$. We will be interested in other values of $c$ below, in the context of the Potts model, so we will treat it as a parameter. For simplicity, we consider the following quasiperiodic potential
 \begin{equation}
\ell_j =-\ln J_j=a+\cosq{j+\theta},\label{eq: initial potential}
\end{equation} 
 with $\varphi$ the golden ratio, $\theta$ is a random phase and $a$ is an arbitrary constant. However, we emphasize that our conclusions also hold for any sufficiently regular quasiperiodic potential with frequency $\varphi$~\cite{supmat}.
  
\begin{figure}[]
\centering
\includegraphics[scale=0.45]{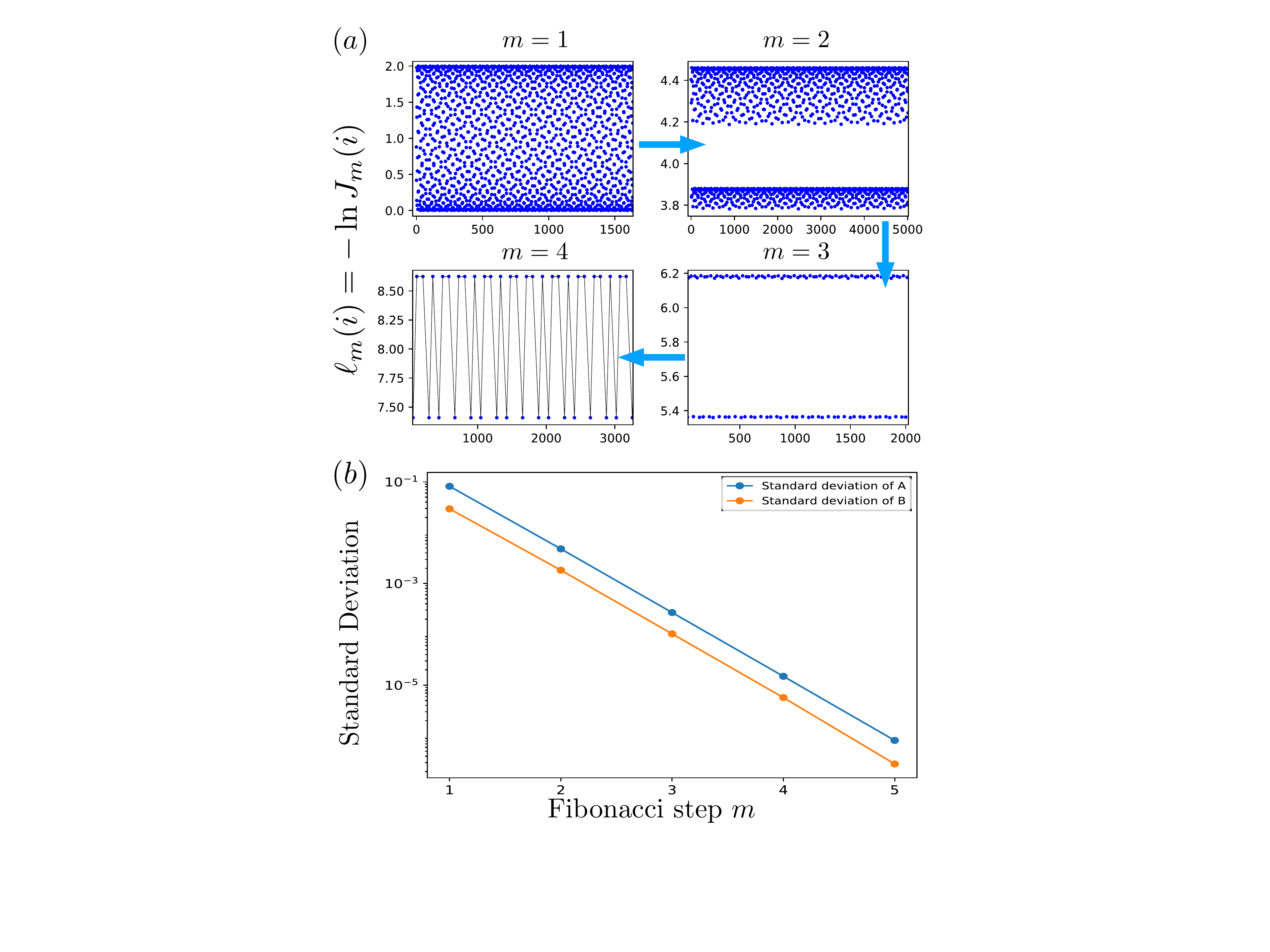}
\caption{{\bf Quasiperiodic Heisenberg chain.} (a) Evolution of the couplings under renormalization for an Heisenberg chain with initial potential~\eqref{eq: initial potential} with $a=1$.
The fluctuations become completely negligible after only 3 Fibonacci steps. (b) The fluctuations about the sequence prediction~\eqref{eq: A_m and B_m for fisher's case}  starting from a cosine potential decay exponentially with the number of Fibonacci RG steps $m$. } \label{Fig: QP to sequence}
\end{figure}

Many properties of the decimation procedure can be understood analytically. It is helpful to introduce the notion of a \emph{local minimum}: i.e., a site $j$ such that $\ell_j < \min(\ell_{j-1}, \ell_{j+1})$. Any such coupling will get decimated before its neighbors, so we can decimate them all at once. It is crucial to note that, if we label minima as $B$ and all other sites as $A$, we will arrive at a two-letter Fibonacci sequence defined by the deflation rule $A\rightarrow AB, B\rightarrow A$~\cite{supmat}. With this observation in mind, we denote the set of minima as $B_0$ and the set of all other couplings as $A_0$. We now decimate all the $B_0$ couplings --- we call this a Fibonacci RG step. This gives rise to a new sequence, in which, once again, we can identify the local minima, denote them $B_1$, decimate them, and so on. Remarkably, after a few steps, we find that all the $A(i)$ and $B(i)$ at a given step become increasingly similar in magnitude. Specifically, 
after $m$ Fibonacci RG steps, we find the effective couplings~\cite{supmat}
\begin{align}
A_m(i)&=a+m(m+1)c+\frac{\cos(F_{3m+2}\pi\varphi )}{\cos{\pi\varphi}}+\epsilon^A_{i,m} ,\nonumber \\
B_m(i)&=a+m^2c-\frac{\cos (F_{3m+1}\pi\varphi )}{\cos{\pi\varphi}} + \epsilon^B_{i,m} , \label{eq: A_m and B_m for fisher's case}
\end{align}
where $F_n$ is the n$^{\rm th}$ Fibonacci number, and $\epsilon^{A,B}_{i,m}$ are site-dependent fluctuations that go to 0 exponentially as $m \to \infty$. With each Fibonacci step the fluctuations get smaller, and we obtain a sharper sequence which asymptotically becomes a perfect binary Fibonacci sequence. Even if the fluctuations are non-negligible in the initial steps of the RG, they are small enough for us to decimate all $B$ couplings at once (Fig. \ref{Fig: QP to sequence}). This means that the complicated initial potential~\eqref{eq: initial potential} flows under RG to a perfect binary Fibonacci sequence. Moreover, we have $A_m-B_m=mc + {o}(1)$ as $m \to \infty$, which indicates that the decimation rule~\eqref{eqRGrule} given by second order perturbation theory becomes asymptotically exact as $m \to \infty$, as in the random case. In contrast with the random case, where the fixed point is specified by a probability distribution of couplings, in the quasiperiodic case the fixed point Hamiltonian is specified by a specific, self-similar sequence of couplings: any sufficiently regular function $f(x) = f(x + \varphi^{-1})$ flows to a binary piecewise function, which is invariant (up to rescaling) under the RG.

We remark that although this flow to nearly discrete sequences is a general property of the RG rules, and occurs for any irrational number, the self-similarity of these discrete sequences is a special property of the Golden Ratio and other irrational numbers with recurring continued fraction expansions, such as the ``metallic means'' $1/(n + 1/(n + \ldots))$ (the Golden Ratio is the case $n = 1$). For more general irrational numbers the sequences do not repeat under the RG: instead, each level of the RG is governed by the coefficient of the continued fraction expansion at the corresponding level~\cite{wilkinson, levitov,PhysRevB.98.134201}.

\emph{Quantum critical behavior}.--- The critical properties of the quasiperiodic chain~\eqref{eqXXX} then follow straightforwardly from eq.~\eqref{eq: A_m and B_m for fisher's case}, in agreement with previous works on  discrete aperiodic sequences. The new fixed point has dynamical exponent $z=\infty$: a chain of length $L \sim \varphi^{3m}$ is fully decimated in $m$ Fibonacci steps, so the energy gap $\Delta E$ of the chain is set by the last coupling to be decimated, $\log \Delta E \sim - m^2 c$, so that $\Delta E \sim e^{-\frac{c}{(3 \ln \varphi)^2} \ln^2 L} $~\cite{hida_new_2004}, where we have used~\eqref{eq: A_m and B_m for fisher's case}. The correlation length exponent $\nu$ can be calculated by introducing an asymmetry between even and odd couplings at the new critical point. 
Let us assume a perfect sequence of $A_0$ and $B_0$ which is dimerized such that $A^E_0=A_0+\delta/2,\ B^E_0=B_0+\delta/2,\ A^O_0=A_0-\delta/2,\ B^O_0=B_0-\delta/2$. 
It is simple to check that if one constructs an effective bond made of $n$ microscopic bonds, its coupling will have an additive piece $\pm n \delta/2$ where the sign depends on whether the effective bond is even or odd. 
Thus, after the RG is iterated out to a scale $L$, the effective couplings at that scale consist of $O(L)$ microscopic couplings, so the asymmetry between the even and odd couplings is $\sim \delta L$. The asymmetry thus becomes of order unity when $L \sim 1/\delta$, and it follows that $\nu = 1$~\cite{supmat}.

\emph{Quantum Potts model}.---We now turn to the $q$-state quantum Potts model, governed by the Hamiltonian
\begin{equation}
H = - \sum\nolimits_i J_i \delta_{n_i, n_{i+1}} - \sum_i  \frac{h_i}{q}  \sum_{n_i,n_i'} |n_i \rangle \langle n'_i |,
\end{equation}
where $n_i$ is a variable on site $i$ that takes one of $q$ possible values. 
To treat this model in the RG scheme, one formally rewrites it as a chain with twice the number of links, and assigns the variables $J_i$ to even links and $h_i$ to odd links. 
The decimation step~\cite{SenthilMajumdar} then takes the same form as Eq.~\eqref{eqRGrule} with $c = \log(q/2)$. When $q > 2$, $c > 0$, so once again the RG flows to discrete sequences. 
The main distinction between the Potts and Heisenberg models lies in the choice of initial couplings: in the Heisenberg model it was natural to draw all bonds from the same quasiperiodic sequence, whereas here it is natural to take the $h_i$ and $J_i$ from distinct quasiperiodic sequences with frequency $\varphi$ but different phases: $J_i, h_i = a + \cos(2 \pi \varphi i + \theta_{J,h}) $, with $a>1$. This introduces a separate variable to the problem, viz. the relative phase $\theta \equiv \theta_{J}-\theta_{h}$ between the sequences for $h_i$ and $J_i$. 

\begin{figure}[]
\centering
\includegraphics[scale=0.37]{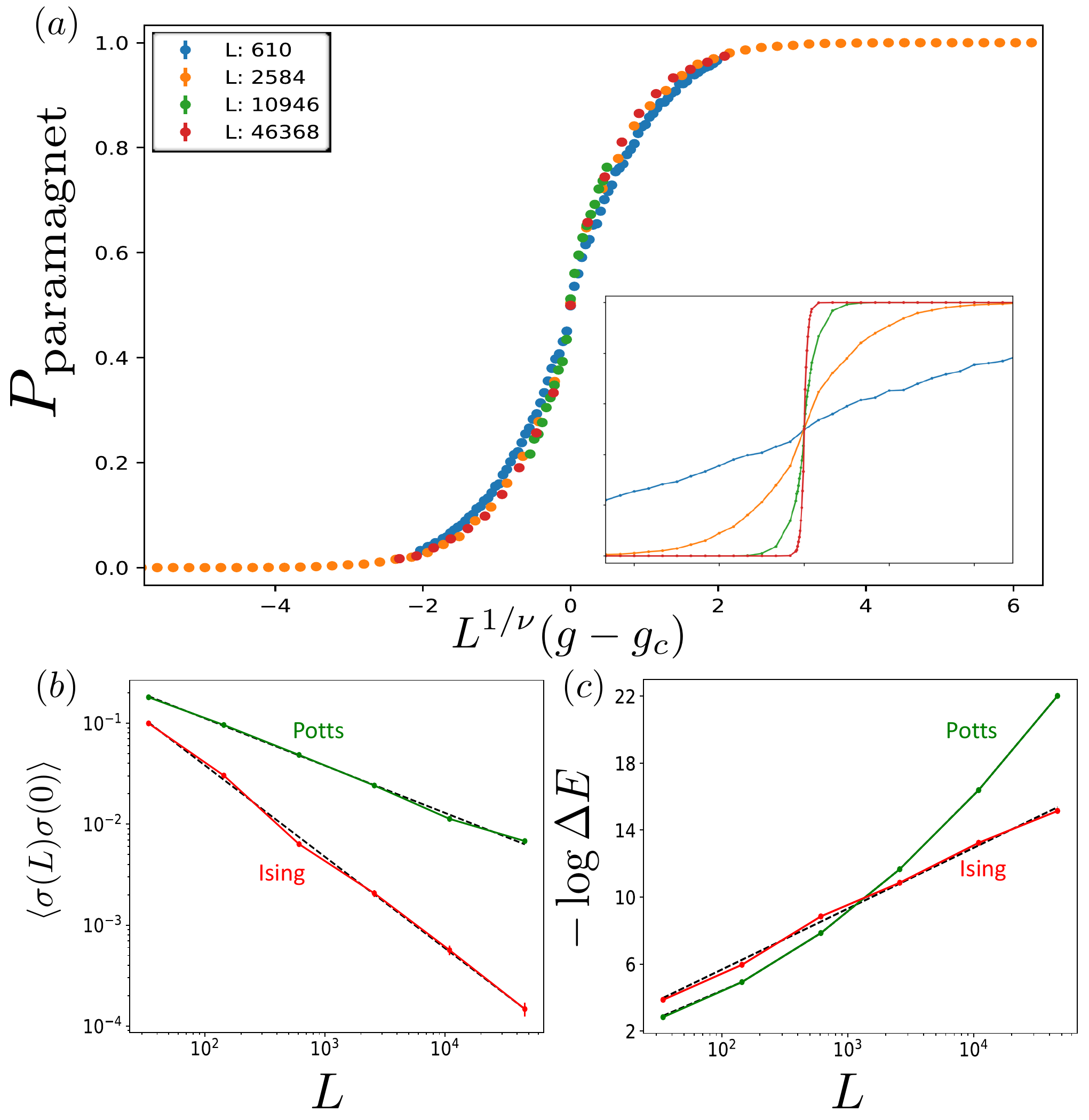}
\caption{{\bf Quasiperiodic Potts ($q=3$) and Ising  ($q=2$)  chains.} For the Ising chain, we choose a potential with both positive and negative couplings, while for the Potts chain, all couplings are taken to be antiferromagnetic. (a) Scaling collapse of the probability of the RG to end in a paramagnetic phase for the Potts model, with $\nu=1$. Here $g$ is an asymmetry parameter between $h_i$ and $J_i$. (Inset) Raw, uncollapsed data. (b) Spin-spin correlation function $\langle \sigma(L) \sigma(0) \rangle $ averaged over the uncorrelated phases $\theta_{J},\theta_{h}$, scaling as $L^{-0.47}$ for Potts (in good agreement with~\eqref{eqSpinSpinPotts} derived for discrete Fibonacci sequences), and $L^{-0.9}$ for Ising. (c) Energy-length scaling: $\Delta E \sim L^{- 0.22 \ln L}$ for Potts, while the Ising transition has a finite dynamical exponent $z \approx 1.6$.  }
 \label{FigPottsIsing}
\end{figure}

Numerically running the RG in this case leads to the following picture. When $\theta$ is close to the special values, we once again observe a flow to self-similar Fibonacci-like sequences. In that case, the results for the XXX spin chain carry over to Potts immediately -- in particular, $\nu=1$. Other critical exponents can readily be computed analytically; for example, we find that correlation function of the order parameter $\sigma^{(a)}_i = \delta_{n_i,a} - q^{-1}$ with $a$ a given Potts color, scales as~\cite{supmat}
\begin{equation}
\langle \sigma^{(a)}_0 \sigma^{(a)}_r \rangle \sim r^{- 2 \Delta_\sigma}, {\rm  with \ }  \Delta_\sigma = \frac{\ln \left(1 + 2 \varphi^{-1} / 3 \right)}{3 \ln \varphi}.
\label{eqSpinSpinPotts}
\end{equation}
However, for large $\theta$ we see abrupt transitions to \emph{different} sequences. Thus, there appear to be \emph{multiple} fixed-point sequences, with ``transitions'' between them occurring at special values of $\theta$. These fixed points have different length-energy scaling: in all cases, $\Delta E \sim e^{- a \ln^2 L}$, but $a$ depends on $\theta$. However, all of the fixed points agree on the exponent $\nu=1$, as well as on the spin-spin correlation function~\eqref{eqSpinSpinPotts} (Fig.~\ref{FigPottsIsing}).

\emph{Ising model}.---We briefly remark on the $q = 2$ Potts model, i.e., the Ising model. In this case, $c = 0$ in the decimation rule~\eqref{eqRGrule}. Thus the RG does not take arbitrary functions to sequences. Instead, nonsingular sequences of $\ell$ are generically squashed under the RG and become effectively constant after a few steps. This corresponds to the observation that weak quasiperiodic modulation is (marginally) irrelevant at the Ising critical point. To see a nontrivial transition in this case, one must take singular distributions of $\ell$; one can do this, e.g., by taking $J_i, h_i = a + \cos(2\pi \varphi i + \theta_{J,h})$, with $0<a<1$ so $\ell_{2 i} = - \ln |J_i | $, $\ell_{2 i+1} = - \ln |h_i | $ is singular. In this model, the critical dynamics is strongly modified because the chain has a finite density of nearly broken links. Although our RG procedure is not controlled for this model, running the RG numerically yields a fixed point that is qualitatively similar to the numerically seen one~\cite{wilkinson, crowley_critical_2018,crowley_quasiperiodic_2018} (Fig.~\ref{FigPottsIsing}): in particular, it has a finite dynamical critical exponent $z \approx 1.6$, which is close to the numerical value $z \approx 2$. The remaining discrepancies are to be expected given that the procedure is not controlled; however, the RG correctly identifies the change in critical behavior that occurs when the distribution of couplings goes through zero.

\emph{Quasiperiodic MBL transition}.---The observation that generic quasiperiodic potentials flow to  analytically-tractable substitution sequences under renormalization has broad implications beyond zero-temperature quantum criticality. In particular, isolated quantum systems can undergo \emph{eigenstate phase transitions}~\cite{Parameswaran_2017}, in which the properties of eigenstates (and correspondingly the nature of the dynamics) changes in a nonanalytic way as some parameter is tuned. A key example of such a transition is the highly studied many-body localization (MBL)~\cite{PhysRev.109.1492, FleishmanAnderson, Altshuler1997QuasiparticleLifetime, Gornyi, BAA, PhysRevB.75.155111, PalHuse, ARCMP,VasseurMoore2016MBLReview,AbaninRev} transition in one dimension, which separates localized and chaotic dynamical phases. In the random case, the nature of the transition remains debated~\cite{PalHuse,abanincriterion,KjallIsing,Luitz,Demler14,PhysRevX.7.021013,khemani_two_2017, PhysRevLett.121.206601}, but many RG schemes have been proposed~\cite{VHA,PVP,DVP,THMdR_meanfield,TMdR_numerics,zhang_many-body_2016,goremykina_analytically_2019}, with the most recent works supporting a Kosterlitz-Thouless-like transition~\cite{goremykina_analytically_2019,PhysRevB.99.094205,PhysRevB.99.224205}.

In comparison with the random case, the physics of MBL in quasiperiodic systems remains largely unexplored. 
Within the latest RG schemes, the mechanism driving the transition is the nucleation of chaos in rare, anomalously thermal regions~\cite{Demler14,mbmott,Agarwal2017rareregion}. 
Quasiperiodic systems seem to lack such regions, and might therefore have a qualitatively different MBL transition, but the nature of this transition is not fully understood (but see Ref.~\cite{PhysRevLett.121.206601}). 
It would be natural to adapt previous RG approaches to the present case; however, there are two inequivalent ways that one could do this. The first approach is to try to construct typical individual eigenstates; in the localized phase, these eigenstates look like random product states. Some RG schemes proceed by locating clusters of spins that are resonant; however, the resonance condition itself depends on the initial product state (e.g., via Hartree effects) and is effectively random. Including these random Hartree shifts would presumably lead to a transition that looks similar to the random one. 
However, an alternative approach is to disentangle the Hamiltonian by a series of unitary rotations, and thus construct localized integrals of motion (LIOMs) in the MBL phase~\cite{Serbyn13-2,Huse13,Serbyn13-1,Imbrie16,ImbriePRL}. 
Deep in the MBL phase, a full set of LIOMs exists; and at a critical disorder value, all (or most) of these LIOMs cease to exist.
The properties of the LIOMs in a quasiperiodic system are clearly quasiperiodic. 
Phenomenological RG schemes that model a system in terms of thermal and localized ``blocks'' can capture this procedure, since they never explicitly invoke eigenstates.

Motivated by this observation, we now explore the simplest such scheme~\cite{zhang_many-body_2016,goremykina_analytically_2019}. This model also describes certain classical coarsening dynamics~\cite{PhysRevE.50.1900}, and as we shall see, its RG rules are similar to those discussed above. 
This model assumes that on some coarse-grained scale, the system can be described alternating thermal (T) and insulating blocks (I), characterized by a single parameter which we call ``length'' $\ell_{T,I}$. During the RG, the smallest I or T block is decimated and merged with its neighbors into a new T/I block, with in the simplest version of this RG~\cite{zhang_many-body_2016}, a renormalized length that is the sum of the decimated blocks $\ell^{T/I}_{\rm new} = \ell^{T/I}_{i-1} +  \ell^{I/T}_{i} +  \ell^{T/I}_{i+1}$. This RG rule is almost the same as~\eqref{eqRGrule}, with $c=0$ and a different sign in front of $\ell_i$. For random initial distributions of the block sizes, this simple model has a second order phase transition with correlation length exponent $\nu \approx 2.5$. Starting from quasiperiodic initial lengths $\ell^{T/I}_i=W^{T/I}(1+\cosq{i+\theta})$, we find that the lengths once again renormalize to a Fibonacci sequence for $W^T = W^I$. 

We now consider perturbing away from these fixed points, e.g., driving the system into the insulating phase by making the $I$ blocks slightly larger than the $T$ blocks. We can do this with the same dimerization perturbation as we previously had. A key thing to note is that this perturbation is actually \emph{irrelevant} at the Fibonacci-sequence fixed point: a small perturbation does not change the sequence of decimations, and the perturbation strength \emph{decreases} under decimation, as a consequence of the crucial sign-change in the RG rule. This situation changes, however, for continuous initial distributions: here, there are pairs of adjacent couplings that are nearby in energy; thus, there are minima (at the critical point) that cease to be minima when the system is perturbed. Thus, even small perturbations change the \emph{order} in which sites are decimated, and thus disrupt the fixed-point sequence. In this sense, dimerization is a dangerous irrelevant perturbation at the quasiperiodic MBL fixed point. 

Interestingly, regardless of this subtlety, we once again find a correlation length exponent $\nu = 1$ for smoothly varying initial conditions. The argument is as follows: define a defect as a bond that was a local minimum for the critical pattern but ceased to be one after being perturbed (or vice versa). If a defect is present at position $i=0$, then another defect will be created at $i=F_n$ with $n$ large enough, since $\varphi F_n$ is almost an integer. Going away from criticality by an amount $\delta W$ induces defects if we choose $n$ so that the deviation of $F_n$ from being the exact period (given by the fractional part of $\varphi F_n$) satifies $\{\varphi F_n\}=\varphi^{-n}\sim\delta W$~\cite{supmat}. The distance between the defects then sets the correlation length $\xi\sim F_n\approx \varphi^n\sim\delta W^{-1}$, implying that $\nu=1$. An interesting consequence is that one can tunably vary $\nu$ by introducing singularities at appropriate points in the original sequence; we have checked that this is in fact true~\cite{supmat}. 

This toy RG is perfectly symmetric between the $T$ and $I$ phases; in realistic models, this symmetry is absent, as thermal blocks are much more ``infectious'' than insulating blocks. To introduce some asymmetry between the two phases, we follow Ref.~\cite{goremykina_analytically_2019}, and introduce a new parameter $\beta$ in the RG rules 
\begin{equation}
\ell^{T/I}_{\rm new} = \ell^{T/I}_{i-1} +   \beta^{I/T}  \ell^{I/T}_{i} + \ell^{T/I}_{i+1}, \label{eqRG_GVS}
\end{equation} 
with $\beta^{I}=1/\beta^T = \beta$. This modified RG is analytically solvable for random distributions, and captures much of the phenomenology of the MBL transition in the maximally asymmetric limit $\beta \to \infty$, with a KT-type transition ($\nu=\infty$). 
In the quasiperiodic case, weak asymmetry is irrelevant at the fixed point; however, as the asymmetry is cranked up, eventually the order of decimations changes, and one flows to a new fixed-point sequence. Thus, as $\beta$ is increased, the system discontinuously jumps between a number of increasingly complicated self-similar fixed-point sequences  which we detail in the Supplemental Material~\cite{supmat}. Perturbations away from criticality lead to defects in these self-similar sequences, with the same exponent $\nu=1$ from the argument above. We therefore conjecture that the value $\nu=1$ remains exact as $\beta \to \infty$. This exponent $\nu=1$ is in agreement with exact diagonalization results~\cite{khemani_two_2017,lee_many-body_2017}, and supports the scenario of two distinct universality classes for the quasiperiodic MBL transition~\cite{khemani_two_2017}. 

\emph{Conclusion}.---To summarize, we have shown that under quite general conditions, conventional real-space RG rules attract generic initial conditions to fixed-point \emph{sequences} that are self-similar under RG. The simplest case is in the Heisenberg spin chain, where the fixed-point sequence is given by the simple Fibonacci deflation rule; many results were previously known for this sequence, but our work shows that these results apply far more generally than had been appreciated. For the Potts model, we again found attracting sequences, but unlike the Heisenberg model we found \emph{multiple} distinct attractors. Finally, for the toy MBL transition models, we found a much more unexpected type of RG flow, in which the control parameter driving the transition is in fact dangerously irrelevant at the critical point, but nevertheless the exponent $\nu = 1$. It would be interesting to investigate whether our results can be extended to two-dimensional quasiperiodic systems~\cite{PhysRevLett.92.047202}.

\emph{Acknowledgments}.--- We thank A. Chandran, P. Crowley, D. Huse, V. Khemani and C. Laumann  for stimulating discussions. We also thank S. Gazit, B.M. Kang and J. Pixley for useful discussions and for sharing with us their unpublished Quantum Monte Carlo results. This work was supported by the US Department of Energy, Office of Science, Basic Energy Sciences, under Early Career Award No. DE-SC0019168 (U.A. and R.V.), the Sloan Foundation through a Sloan Research Fellowship (R.V.), and by NSF Grant No. DMR-1653271 (S.G.).

\bibliography{References}

%

%

\bigskip

\onecolumngrid
\newpage



\section*{Supplementary material}

# Supplemental Material for "Universality and Quantum Criticality in Quasi-periodic Spin Chains"

Utkarsh Agrawal,[1] Sarang Gopalakrishnan,[2] and Romain Vasseur[1]

[1] *Department of Physics, University of Massachusetts, Amherst, MA 01003, USA*
[2] *Department of Physics and Astronomy, CUNY College of Staten Island,*
*Staten Island, NY 10314; Physics Program and Initiative for the Theoretical Sciences,*
*The Graduate Center, CUNY, New York, NY 10016, USA*

(Dated: August 5, 2019)

## CONTENTS

## I. GROUND STATES OF QUASIPERIODIC SPIN CHAINS

This section elaborates on various aspects of the ground-state physics of quasiperiodic spin chains. The key result of this section is an explanation of why generic quasiperiodic patterns flow to sequences; we also discuss other aspects of quantum criticality in these spin chains.

### A. Pattern of the local minima in the initial potential

We noted in the main text that the local minima of the initial potential $1 + \cos(2\pi\varphi i + \theta)$—i.e., sites at which $\ell$ is smaller than at either of the neighbors—follow a sequence structure that sharpens into a discrete sequence under renormalization. We here provide a proof of this statement for this specific potential. It is helpful, for this section, to define the *fractional part* of a real number $x$, denoted $\{x\}$, as the difference between $x$ and the largest integer less than $x$, i.e., $\{x\} \equiv x - \lfloor x \rfloor$.

We start by analyzing the pattern of minima in the initial potential. Let us suppose that for a given coupling $n$, $\ell_n = 1 + \cos 2\pi\varphi n$ is a local minimum (we take $\theta = 0$ for now). By definition, we have $\ell_{n+1} > \ell_n$. Denote the corresponding values of the position $i$ as $x$ and $y$. There are two cases: $\{x\} < 1/2$ and $\{x\} > 1/2$. When $\{x\} < 1/2$, then the fact that $\ell_{n+1} > \ell_n$ implies that $\{y\} \in (0, \{x\}) \cup (1 - \{x\}, 1)$, whereas if $\{x\} > 1/2$ then $\{y\} \in (0, 1 - \{x\}) \cup (\{x\}, 1)$. From the definition of the potential, we have $y = (n+1)\phi$ and $x = n\phi$. Studying various cases of $0 < \{n\varphi\} < 1 - \{\varphi\}$, $1 - \{\varphi\} < \{n\varphi\} < 1/2$ and $\{n\varphi\} > 1/2$, we get $\frac{1-\{\varphi\}}{2} < \{n\varphi\} < 1 - \frac{\{\varphi\}}{2}$. Applying the same logic to the inequality $\ell_{n-1} > \ell_n$, we find $\frac{\{\varphi\}}{2} < \{n\varphi\} < \frac{1+\{\varphi\}}{2}$. Taking the intersection of these two inequalities, the condition for $\ell_n$ to be a local minimum reads

$$\frac{\{\varphi\}}{2} < \{n\varphi\} < 1 - \frac{\{\varphi\}}{2}.$$

Reintroducing the phase $\theta$ to the initial potential modifies the above result to $\frac{\{\varphi\}}{2} < \{n\varphi + \frac{\theta}{2\pi}\} < 1 - \frac{\{\varphi\}}{2}$. This can be written as,

$$\boxed{0 < \{n\varphi + \frac{\theta}{2\pi} - \frac{\{\varphi\}}{2}\} < 1 - \{\varphi\}.}$$

We now identify this pattern with the Fibonacci binary sequence defined by the inflation rules $A \to AB$, $B \to A$. Recall that the inflation rule is equivalent to stating that site $n$ is a B site iff $\lfloor(n+n_0+1)\varphi\rfloor - \lfloor(n+n_0)\varphi\rfloor = 1$, where the sequence—starting from the initial letter A, then applying the inflation rules—begins at site $n_0$. Here, $\lfloor \cdot \rfloor$ denotes the largest integer smaller than $\cdot$. Thus, $n_0$ merely translates the standard Fibonacci sequence. Writing $\lfloor(n+n_0+1)\varphi\rfloor$ as $\lfloor(n+n_0)\varphi\rfloor + \lfloor\varphi\rfloor + \lfloor\{(n+n_0)\varphi\} + \{\varphi\}\rfloor$ yields $\lfloor\{(n+n_0)\varphi\} + \{\varphi\}\rfloor = 0$. This is true if and only if $0 < \{n\varphi + n_0\varphi\} < 1 - \{\varphi\}$.

Thus we have the following chain of results:

$$(\ell_n \text{ is a local minimum}) \iff 0 < \left\{n\varphi + \frac{\theta}{2\pi} - \frac{\{\varphi\}}{2}\right\} < 1 - \{\varphi\} \iff$$

$$\Big(\text{letter '}B\text{' is at } n^{\text{th}} \text{ position in the Fibonacci word sequence shifted from}$$

$$\text{the standard Fibonacci sequence by distance } n_0 \text{ such that } \{n_0\varphi\} = \{\frac{\theta}{2\pi} - \frac{\{\varphi\}}{2}\}\Big),$$

proving the claim that local minima of the initial potential indeed follow the pattern of the letter 'B' in the Fibonacci sequence.

Thanks to the above result, we can decompose the cosine coupling into A-couplings and B-couplings arranged in a Fibonacci sequence. To emphasize this decomposition, we shall relabel the couplings $\ell_n$ as follows:

$$\ell_n \equiv \begin{cases} A_0(n), & \ell_n \text{ is not a minimum} \\ B_0(n), & \ell_n \text{ is a minimum} \end{cases} \tag{1}$$

Note that $A_0(n)$ and $B_0(n)$ are not defined for all values of $n$: for a given value of $n$, either $A_0(n)$ or $B_0(n)$ is defined.

### B. Sharpening of the sequence structure under RG

For the XXX spin chain, the RG rules are given by

$$\ell_{\text{eff}} = \ell_{n+1} - \ell_n + \ell_{n-1} + c, \tag{2}$$

with $c = \ln 2$ where $\ell_n$ is the smallest $\ell$. We remark that the whole Fibonacci sequence is built up of word patterns 'ABABA' and 'ABA'. This means that we can apply the decimate all the local minima $B$ in one go, arriving at the following renormalized couplings:

$$A_{m+1}(n) = A_m(n-2) - B_m(n-1) + A_m(n) - B_m(n+1) + A_m(n+2) + 2c$$
$$B_{m+1}(n) = A_m(n-1) - B_m(n) + A_m(n+1) + c,$$

where $m$ labels the number of such Fibonacci RG steps. In order to get $A_{m+1}$ we decimate two $B_m$'s, and to get $B_{m+1}$ we decimate one $B_m$. If $A_m$ and $B_m$ form a Fibonacci sequence then $A_{m+1}$ and $B_{m+1}$ also follow the Fibonacci pattern. This follows from the inflation rule: $A \to ABABA$ and $B \to ABA$. Thus starting from a Fibonacci word sequence, if we replace 'ABABA' by 'A' and 'ABA' by 'B', we again get a new Fibonacci sequence.

Note: We are considering periodic boundary conditions, so our initial system size should be an even number. Also we will consider system sizes given by Fibonacci numbers, $F_n$, as the fractional part of $F_n\varphi$ is $\varphi^{-n}$, i.e. $F_n\varphi \approx F_{n+1}$. Since all even Fibonacci numbers can be written as $F_{3l}$, we consider initial system sizes of the form $N = F_{3l}$. In a Fibonacci sequence of size $F_{3l}$, we have $F_{3l-2}$ $B$ letters and $F_{3l-1}$ $A$ letters. Since this is a word sequence we can decimate all the $F_{3l-1}$ 'B' couplings in one go. Each decimation decreases the number of couplings (bonds) by 2. So after $F_{3l-2}$ decimations, the number of remaining bonds is $F_{3l} - 2F_{3l-2} = F_{3(l-1)}$. This corresponds to what we call a Fibonacci step.

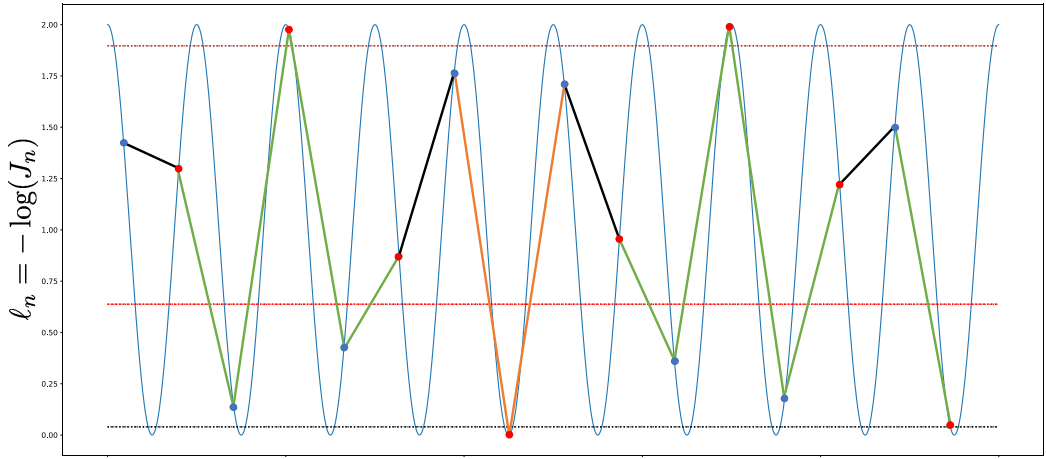

FIG. 1. Part of the initial coupling distribution. The green 'W' shape is the pattern 'ABABA' in the Fibonacci sequence; this will give rise to $A$ block in the next generation of Fibonacci sequence. The orange 'V' shape is the pattern 'ABA' which will lead to a $B$ block in the next generation. Any block with length in the region above the brown dashed line will be the middle 'A' block in the pattern 'ABABA', while blocks in the region between the red and blue dashed line will constitute the 'B' blocks in 'ABABA'. Blocks with length below the blue dashed line will be a 'B' block in the pattern 'ABA'.

After the first Fibonacci step, we have two bands of couplings, A and B, given by,

$$A_1(n) = \sum_{i=-2,0,2}(1 + \cos(2\pi\varphi(n+i))) - \sum_{i=-1,1}(1 + \cos(1 + 2\pi\varphi(n+i))) + 2c,$$
$$= 1 + 2c + \cos(2\pi\varphi n)(1 + 2(\cos(4\pi\varphi) - \cos(2\pi\varphi))),$$
$$= 1 + 2c + \lambda_1\cos(2\pi\varphi n).$$
$$B_1(n) = \sum_{i=-1,1}(1 + \cos(1 + 2\pi\varphi(n+i))) - (1 + \cos(2\pi\varphi(n))) + c,$$
$$= 1 + c - \cos(2\pi\varphi n)(1 - 2\cos(2\pi\varphi)),$$
$$= 1 + c + \lambda_2\cos(2\pi\varphi n),$$

where $(1 + 2(\cos(4\pi\varphi) - \cos(2\pi\varphi))) \equiv \lambda_1 \approx 2.64$ and $-(1 - 2\cos(2\pi\varphi)) \equiv \lambda_2 \approx -2.47$. The label $n$ in $A_1(n)$ and $B_1(n)$ only takes certain integer values, corresponding respectively to the positions—on the original lattice—of the central bonds on the $ABABA$ and $ABA$ clusters. Thus for instance (Fig. 1) if $A_1(0)$ is defined, at the center of the $A$ cluster, then $B_1(n)$ can only be defined for $n = 4$. One can straightforwardly check that at this level all $B_1(n)$ that are defined are local minima, i.e., they are smaller than the neighboring $A_1$ couplings.

By considering the range of possible values of $n$ [i.e., the on-site phase] for which an ABABA sequence is possible, we find the following constraint:

$$1 - \cos(2\pi\varphi n) < \delta A_1, \quad \delta A_1 = 1 - \cos\left(2\pi\frac{-3\varphi+5}{2}\right)[\approx 0.1]. \tag{3}$$

Thus the bandwidth for $A$-type couplings on the new, decimated chain is $\approx 0.1$. A similar analysis for the B couplings—now considering patterns of the form AABAA—yields the constraint $\cos(2\pi\varphi n) \in (-1, -1 + \delta B_1)$ where $\delta B_1 = 1 - \cos\left(2\pi\frac{5\varphi-8}{2}\right)(\approx 0.04)$. See Fig. 1. To summarize, after one step of decimation, all the couplings within each "type" (A or B) become similar in value, but are separated from couplings of the other type by $\approx c$. This leads to the formation of well-separated bands, which then become increasingly well-separated under the action of the RG.

To form $A$ and $B$ couplings in the subsequent Fibonacci steps, the restriction on the initial value of the middle couplings becomes tighter. For example, to get an $A_2$ coupling we need to have 'ABABA$ABA$ABABA$ABA$ABABA' patterns in the initial potential, but to have such a long pattern the middle 'A' coupling needs to be closer to $\cos 2\pi\varphi n = 1$ than it was required to form $A_1$. Thus we expect the fluctuations $\delta A_m$ and $\delta B_m$ to go to zero as $m \to \infty$. Empirically these fluctuations decay exponentially with $m$.

This can be iterated to get (we are using $C(n) \equiv \cos 2\pi\varphi n$ and $S(n) \equiv \sin 2\pi\varphi n$ for brevity),

$$A_m(n) = \sum_{i=-\frac{F_{3m+2}-1}{2}}^{\frac{F_{3m+2}-1}{2}} (-1)^i \left(1 + \cos 2\pi\varphi(n+i)\right) + m(m+1)c,$$

$$= 1 + m(m+1)c + C(n)\left(1 + 2\frac{S(\frac{F_{3m+2}-1}{4})}{S(1)}\left(C\left(\frac{F_{3m+2}+3}{4}\right) - C\left(\frac{F_{3m+2}-1}{4}\right)\right)\right),$$

$$= 1 + m(m+1)c + C(n)\left(1 - 2\frac{S(\frac{F_{3m+2}-1}{4})S(\frac{F_{3m+2}+1}{4})}{C(1/2)}\right),$$

$$= 1 + m(m+1)c + C(n)\left(1 - \frac{C(1/2) - C(\frac{F_{3m+2}}{2})}{C(1/2)}\right),$$

$$= 1 + m(m+1)c + \frac{C(\frac{F_{3m+2}}{2})}{C(1/2)}C(n).$$

Similarly we have

$$B_m(n) = 1 + m^2c - (-1)^m C(n)\frac{C(\frac{F_{3m+1}-3}{4\varphi} + \frac{F_{3m+1}}{2})}{C(1/2)},$$

$$= 1 + m^2c - (-1)^m C(n)\frac{(-1)^{m+1}C(\frac{F_{3m+1}}{2})}{C(1/2)},$$

$$= 1 + m^2c + \frac{C(\frac{F_{3m+1}}{2})}{C(1/2)}C(n).$$

For large m, $A_m(n) \approx 1 + m(m+1)c + \frac{1}{\cos\pi\varphi}$ and $B_m(n) \approx 1 + m^2c + \frac{1}{\cos\pi\varphi}$. $A_m - B_m \xrightarrow{m\to\infty} mc$. If $c = 0$ (corresponding physically to Ising and XX chains that can be mapped onto free fermions), then the sequence structure does not survive under RG.

*Generic potentials:* We showed above that for couplings of the form $-\log(J_n) = a + \cos(2\pi\varphi n + \theta)$, the initial distribution flows to a sequence under RG. But this result is quite general and almost all initial distribution of the couplings flow to a sequence. If $f$ is any monotonic function in the range $(-1, 1)$, is bounded and $\ell_n = -\log J_n = f(\cos(2\pi\varphi n + \theta))$, then the minima can be identified exactly as in the analysis above. Therefore, under RG, the couplings flow to the Fibonacci sequence. Natural functions like logarithm, exponential, etc. all satisfy the above criteria. For example, couplings of the form $\ell_n = -\log(1 + \epsilon + \cos(2\pi\varphi n + \theta))$ (corresponding to the natural choice $J_n = 1 + \epsilon + \cos(2\pi\varphi n + \theta)$) will flow to a sequence – and we verified this numerically – where $\epsilon$ is a positive constant which prevents the couplings from hitting the singularity. We have also checked that adding higher harmonics to $J_n$ such as $\cos(4\pi\varphi n + \delta)$ with a different phase $\delta$ does not affect the flow to sequences.

Note that $f$ need not be a monotonic function. E.g for $f(x) = |0.5 + x|$ and $f(x) = \sin(2x)$, we get Fibonacci sequences under RG flow even though the above functions are non-monotonic in the range (-1,1). But if the function $f$ has too many extrema, we find that the Fibonacci sequence structure gets destroyed, e.g $f(x) = \sin(4x)$.

## C. Correlation length exponent

The fixed points of the above RG is unstable against dimerization of even and odd couplings. To find the critical exponent $\nu$ associated with the fixed point, we study the behavior of the sequence under an asymmetric perturbation. We dimerize the perfect sequence as $A_0^E = A_0 + \delta^E$, $B_0^E = B_0 + \delta^E$, $A_0^O = A_0 - \delta^O$, $B_0^O = B_0 - \delta^O$. Under RG, after $m$ Fibonacci steps the sequence flows to,

$$A_m^{E,O} = (2m+1)A_0 - 2mB_0 + m(m+1)c \pm \frac{F_{3m+2}+1}{2}\delta^{E,O} \pm \frac{F_{3m+2}-1}{2}\delta^{O,E}$$

$$B_m^{E,O} = 2mA_0 - (2m-1)B_0 + m^2c \pm \frac{F_{3m+1}+1}{2}\delta^{E,O} \pm \frac{F_{3m+1}-1}{2}\delta^{O,E}.$$

The asymmetry between even and odd blocks keeps increasing with $m$ and eventually $B^E - A^O = B_0 - A_0 - mc + \frac{F_{3m+3}}{2}(\delta^E + \delta^O)$ will become positive, destroying the Fibonacci pattern and driving the system to a phase. Taking $\delta^E = \delta^O = \delta_0$, the dimerization parameter $\delta_m = A_m^E + B_m^E - (A_m^O + B_m^O) = F_{3m+3}\delta_0/2$ scales under RG as $\mathcal{R}[\delta_m] \equiv \delta_{m+1} \approx \varphi^3 \delta_m$ with the RG eigenvalue $\lambda_\delta$ given by $\lambda_\delta = \varphi^3 \equiv b^{y_\delta}$. In a single Fibonacci step, the system size is scaled by $\varphi^3$ implying that $b = \varphi^3$. This gives us $\nu \equiv 1/y_\delta = 1$.

### D. Spin exponent for the quasiperiodic quantum Potts chain

At the critical point, the 2-point correlation function of the Potts order parameter scales as,

$$C(0,r) = \langle \sigma_0 \sigma_r \rangle \sim \frac{1}{r^{2\Delta_\sigma}}. \tag{4}$$

Assuming for now that the critical point of the quantum Potts model is given by the Fibonacci fixed point discussed above, we can calculate this 2-point function analytically. Decimating the $J_i$ couplings in the RG leads to the formation of spin clusters which are locally ferromagnetic, while the decimations of a transverse field term $h_i$ freezes the corresponding spin (or effective spin cluster) in a given configuration with zero magnetization. We call the spin clusters that are not yet decimated active. The 2-point function is equal to the probability of the spins at 0 and $r$ to be part the same active cluster, which is true if and only if all the spins between them are either part of the same cluster to which the spins at 0 and $r$ will eventually belong to, or are decimated. This implies that the probability for the two spins to be in same cluster is equal to the probability for them not to be decimated at the length scale where there is no other active cluster in between them. If we denote by $P(r)$ the probability for a given spin not be decimated under RG over a distance $r$, we thus have $C(r) \sim P(r)^2$.

We then compute $P(r) \sim r^{-\Delta_\sigma}$ as follows: we start with system size $r = F_{3m}$ with $h_i$ and $J_i$ taken from the same Fibonacci sequence, with $h$ corresponding to odd links and $J$ even ones. Let $h_0$ denote the magnetic field associated with the cluster under consideration. For a Fibonacci sequence, we know that all 'B' couplings are decimated in a single Fibonacci step. Thus we have following relation,

$$\begin{aligned}
P(r) &= (\text{Prob that } h_0 \text{ is not decimated}) \\
&= \prod_{i=0}^{m} P(h_0 \text{ is a 'A' type bond after } i \text{ Fibonacci steps}) \\
&\equiv \prod_{i=0}^{m} P_i.
\end{aligned} \tag{5}$$

$P_0$ represents the probability that the relevant field $h_0$ is a type 'A' coupling in the initial distribution, and is given by $P_0 = \varphi/(1 + \varphi)$. ($P_0$ is the probability of getting a 'A' coupling at some particular site for a randomly chosen global phase.) $P_1$ is the probability that the $h_0$ coupling is a type 'A' link after 1 Fibonacci step. $P_1$ thus denotes the probability that the bond in question is part of a sub-pattern in previous Fibonacci step which gives birth to a 'A' bond in this step. For the standard Fibonacci sequence we know this sub-pattern to be 'ABABA'. Thus $P_1 = $ probability of the link to be in the sub-pattern 'ABABA' in the previous Fibonacci step $= 3\varphi/(3\varphi + 2)$. In a similar fashion we can show that all remaining $P_i = 3\varphi/(3\varphi + 2)$. This leads to,

$$P(r) = \frac{\varphi}{\varphi + 1}\left(\frac{3\varphi}{3\varphi + 2}\right)^{m-1} \sim \left(\frac{3\varphi}{3\varphi + 2}\right)^{\ln r/3\ln \varphi} \sim r^{-\Delta_\sigma},$$

with $\Delta_\sigma = \ln\left(1 + 2\varphi^{-1}/3\right)/\left(3\ln \varphi\right)$.

For a quasiperiodic Potts chain, it is much more natural to take $h$ and $J$ fields from independent distinct quasiperiodic potentials. The above analysis for the Fibonacci fixed point will not apply for such initial distribution. The system, however, still flows to sequences, but the fixed point oscillates between multiple sequences. Though the details of these sequences differ, we observe numerically that they have same universal features which leads to same spin and correlation length exponent as in above calculations (see main text). We remark that our argument only used the probabilities/ratio of various sub-patterns and letters, which could be same for a wide class of sequences. The details of the sequence is unimportant. As an example, for many choices of global phase we get a 3 letter sequence defined by the inflation rules $A_1 \rightarrow A_1 B A_2 B A_1$, $A_2 \rightarrow A_2 B A_1 B A_2$, $B \rightarrow A_1 B A_2$. This sequence has same ratios and probabilities of relevant sub-patterns as the Fibonacci sequence and hence the above calculations carry forward.

## II. QUASI-PERIODIC MBL TRANSITION

### A. Symmetric MBL RG ($\beta^I = 1$)

#### 1. Flow to Fibonacci Sequence

The arguments above can be readily generalized to the symmetric MBL RG (see main text). The RG assumes alternating thermal and insulating blocks parametrized by "lengths", $\ell^{T/I}$. The RG flow is dictated by the decimation of the lowest length to get a new length given by the rule,

$$\ell_{\text{new}}^{T/I} = \ell_{n-1}^{T/I} + \beta^{I/T}\ell_n^{I/T} + \ell_{n+1}^{T/I}, \tag{6}$$

where $\beta^I = 1/\beta^T \geq 1$. We take the initial distribution of lengths to be a quasi-periodic potential given by $\ell_n^{T/I} = W^{T/I}(1 + \cos(2\pi\varphi n + \theta))$. The initial length distribution can be thought of as a sequence like in the case of XXX spin chain, and for the symmetric case of $\beta^I = 1$ we get a Fibonacci sequence under RG with the following values,

$$A_m(n) = \sum_{i=-\frac{F_{3m+2}-1}{2}}^{i=\frac{F_{3m+2}-1}{2}} (1 + \cos(2\pi\varphi(n-i))) \tag{7}$$

$$= F_{3m+2} + \cos(2\pi\varphi n)\left(1 + 2\frac{\cos\left(\pi\varphi\frac{F_{3m+2}+1}{2}\right)\sin\left(\pi\varphi\frac{F_{3m+2}-1}{2}\right)}{\sin(\pi\varphi)}\right)$$

$$= F_{3m+2} + \left(1 + \frac{\sin(F_{3m+3}\pi\varphi) - \sin(\pi\varphi)}{\sin(\pi\varphi)}\right) \times \cos(2\pi\varphi n)$$

$$= F_{3m+2} + \frac{\sin(F_{3m+3}\pi\varphi)}{\sin(\pi\varphi)}\cos(2\pi\varphi n)$$

$$\xrightarrow{m\to\infty} F_{3m+2},$$

since $\sin(F_{3m+3}\pi\varphi) = 0$ as $F_{3m+3}\varphi \approx F_{3m+4}$. Similarly one can show that,

$$B_m(n) = F_{3m+1} + \frac{\sin(F_{3m+1}\pi\varphi)}{\sin(\pi\varphi)}\cos(2\pi\varphi n)$$

$$\xrightarrow{m\to\infty} F_{3m+1} \tag{8}$$

Notice that $A_m - B_m = F_{3m+2} - F_{3m+1} \approx \varphi^{3m+1}(\varphi - 1)$, implying that under RG the initial length distribution flows to a fixed point characterized by infinite difference in length. Thus the assumption that $\ell_n \ll \ell_{n\pm1}$ gets better with the RG flow.

As we increase $\beta^I$ slightly, the thermal blocks are "favored" by the RG rules. This changes (7) and (8) slightly to give (ignoring fluctuations),

$$A_m^T = F_{3m+2} + \delta_{1,m}^T,$$
$$A_m^I = F_{3m+2} - \delta_{1,m}^I,$$
$$B_m^T = F_{3m+1} + \delta_{2,m}^T,$$
$$B_m^I = F_{3m+1} - \delta_{2,m}^I,$$

where $\delta_{1,2}^{T,I} > 0$, and represents the fact that the size of insulating blocks are reduced due to $\beta^T < 1$ and the size of thermal blocks are increased due to $\beta^I > 1$. Under RG we will have,

$$A_{m+1}^T = 3A_m^T + 2\beta^I B_m^I = F_{3(m+1)+2} + 3\delta_{1,m}^T - 2(\beta^I - 1)\delta_{2,m}^I,$$
$$A_{m+1}^I = 3A_m^I + 2\beta_T B_m^T = F_{3(m+1)+2} - 3\delta_{1,m}^I - 2(1 - \beta_T)\delta_{2,m}^T,$$
$$B_m^T = 2A_m^T + \beta^I B_m^I = F_{3(m+1)+1} + 2\delta_{1,m}^T - (\beta^I - 1)\delta_{2,m}^I,$$
$$B_m^I = 2A_m^I + \beta_T B_m^T = F_{3(m+1)+1} - 2\delta_{1,m}^I - (1 - \beta_T)\delta_{2,m}^T.$$

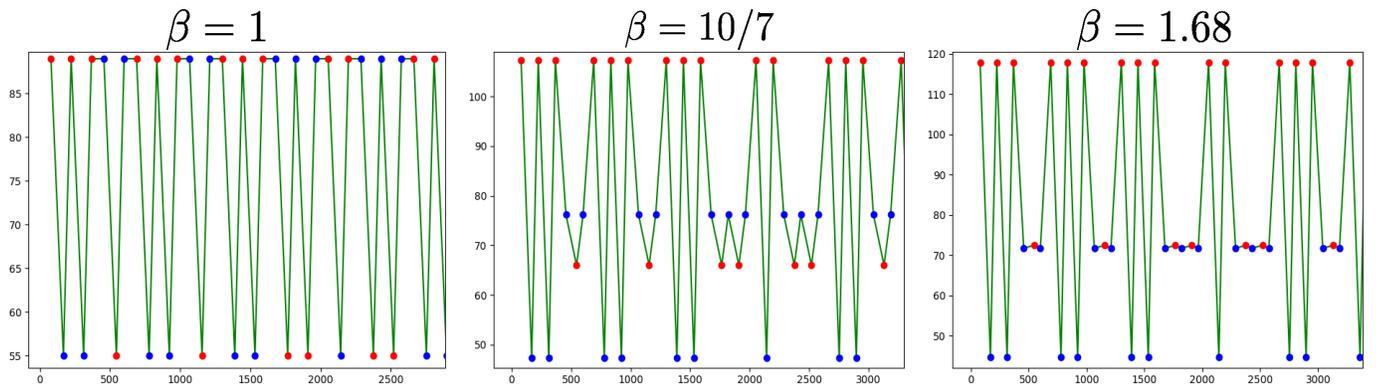

FIG. 2. Critical self-similar sequence for the MBL transition RG showing the lengths $\ell_n$ after 3 Fibonacci step for different values of $\beta^I$, and $\delta W = 0$. The red (blue) dots represent the thermal (insulating) blocks. We see that on increasing $\beta^I$ slightly, the length of thermal blocks become greater but the the sequence structure is not broken. However, when $\beta^I$ is increased further, low thermal blocks become greater than high insulating blocks (see $\beta^I = 1.68$).

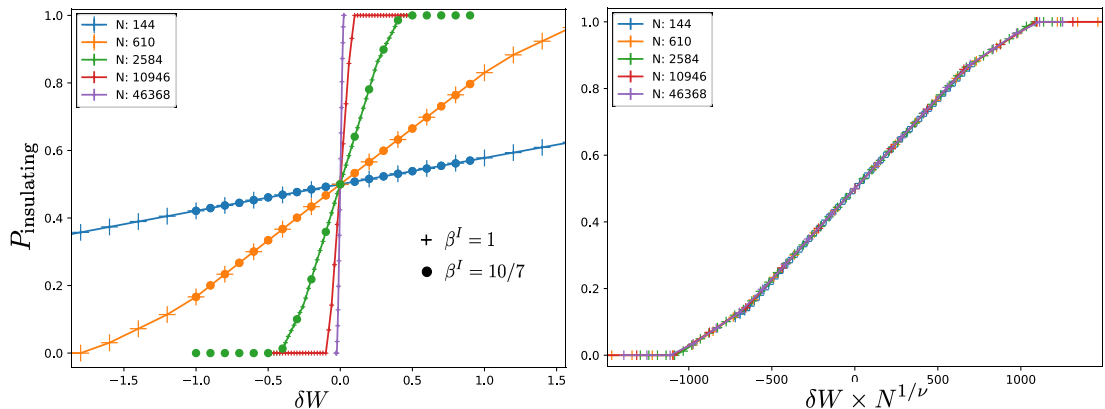

FIG. 3. *Left:* Probability of getting an insulating block at the end of RG is plotted against $\delta W \equiv W^I - W^T$ for $N$ initial blocks. The '+' markers correspond to $\beta_T = \beta^I = 1$ (symmetric case), while the 'o' markers correspond to $\beta^I = 10/7$. The data for both $\beta^I$ overlap, as expected from the arguments given in the text. *Right:* Finite size collapse for $\beta^I = 1$ with $\nu = 1$.

The first term in the above expression increases by a factor of $\varphi^3$ and the correction due to the asymmetry between thermal and insulating blocks also increases by an almost same factor. Thus if initially the corrections are small enough we see that the sequence structure is not broken under the RG flow. But for a large asymmetry, $\delta_{1,2}$ can be large enough to destroy the sequence pattern (as we get $B^T > A^I$). In fact studying the dependence of $\delta_{1,2}^{T,I}$ on $\beta^I$, we find that for $1 < \beta^I < 1.6$ we have the same Fibonacci self-similar sequence as in symmetric case. This can be confirmed numerically by running the RG for different values of $\beta^I$. In the left pannel of Fig. 3 we compare transition data for $\beta^I = 1$ and $\beta^I = 10/7$. They are overlapping perfectly in agreement with our argument.

### 2. Correlation length exponent

We now compute the critical exponent $\nu$ in the symmetric case ($\beta^I = 1$). The mechanism responsible for driving the system to a phase (either thermal or MBL) is by introducing an asymmetry in the initial distribution via $W^I \neq W^T$. This leads to formation of "defects", see Fig. 4.

Let us assume that $W^T = 1$ and $\delta \equiv (W^I - 1) > 0$ is small, so that the change in insulating blocks close to the red dash line in Fig. 4 can be written as, $W^I(1 + \cos(2\pi(\varphi n + \theta)) \equiv 1 + \cos(2\pi(\varphi n + \Delta) + \theta)$, where $\Delta$ is a function of $\delta W$ with $\Delta \propto \delta W$ for small $\delta W$ (provided the derivative of the initial potential is well defined at that point), see Fig. 4 and 5. The phase plays no important role so we take $\theta = 0$ for simplicity. At criticality ($\delta W = 0$) and at the boundary of the minima region, i.e on the dashed red line in Fig. 4, two adjacent insulating and thermal blocks have

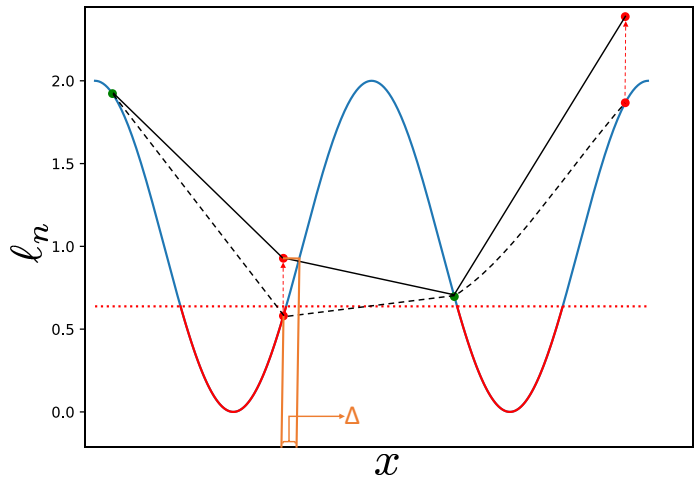

FIG. 4. Formation of a defect in the sequence. The green dots represent thermal blocks and red dots are insulating. The red dashed line represents the boundary for a block to be a local minimum: all the local minima are below that line for $W^T = W^I = 1$. Here 4 consecutive blocks are shown. As we increase $W^I$, the thermal (red) dots go up. The initial potential (dashed black line) gets transformed into black line, showing the shift of the minima from the red dot to the green dot.

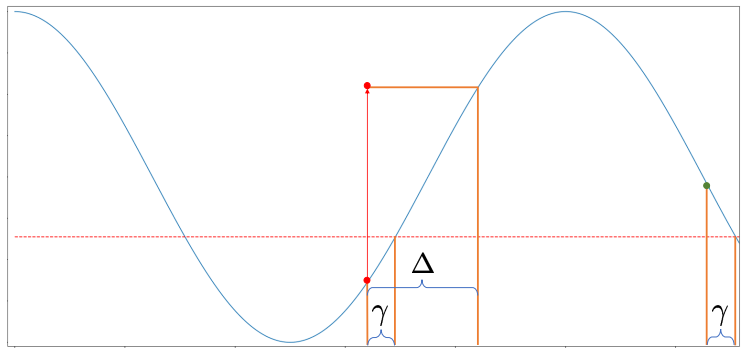

FIG. 5. The red dot is a local minimum insulating block while the green dot corresponds to the adjacent thermal block. The red dashed line has the same meaning as in Fig. 4: all local minima of the initial potential are below this line. As one introduces an asymmetry $\delta W$, the red dot goes above the green dot (corresponding to the formation of a defect) if and only if $\Delta - \gamma > \gamma$.

the same length, that is, if $\ell_n$ lies on the red line then $\{n\varphi\} = 1 - \{\varphi\}/2$ and $\{(n+1)\varphi\} = \{\varphi\}/2$. Now, imagine a local minimum insulating block slightly below the dashed line, i.e $\{n\varphi\} = 1 - \{\varphi\}/2 - \gamma$, then the adjacent thermal block will be above the red line by same amount with $\{(n+1)\varphi\} = \{\varphi\}/2 - \gamma$. As we move away from criticality $\delta W \neq 0$, we have (see Fig. 5)

$$\{n\varphi\} \xrightarrow{\delta W \neq 0} 1 - \{\varphi\}/2 - \gamma + \Delta.$$

Now if, $-\gamma + \Delta > \gamma$ then the insulating block will become larger than the neighboring thermal block. This implies that any insulating block for which

$$1 - \{\varphi\}/2 - \Delta/2 < \{n\varphi\} < 1 - \{\varphi\}/2$$

leads to a defect. See Fig. 5.

To compute the correlation length, suppose that the block at $n_0$ is arbitrarily close to the red line and satisfies the above criterion for creating a defect. We are now interested in the position of the next defect in the sequence pattern. Let $n_0 + k$ be the position of the next defect. We want $k$ to satisfy $-\Delta/2 < k\varphi - l < 0$ for some integer $l$. Since $n_0 + k$ is the first defect after $n_0$, we should also impose the constraint that $0 < p\varphi - q < 1/2$ or $-1/2 < p\varphi - q < -\Delta/2$, $\forall p < k$ and for some $q \in \mathbf{Z}$. We know by Diophantine approximation that the quantity $|a\varphi - b|$, where $a, b$ are integers, is lowest among all the integers $c < a$ *if and only if* $a$ is a Fibonacci number. Thus $k$ has to be a Fibonacci number,

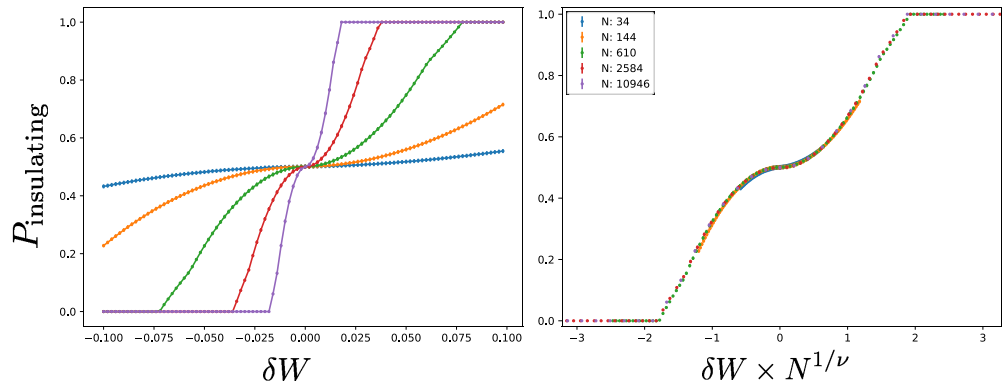

FIG. 6. Transition and scaling collapse for the symmetric) MBL transition RG ($\beta^I = 1$) for a singular initial distribution $\ell_n = 2 + \text{sgn}\left(\cos(2\pi\frac{\{\varphi\}}{2}) - \cos(2\pi\varphi n + \theta)\right)\sqrt{|\cos(2\pi\frac{\{\varphi\}}{2}) - \cos(2\pi\varphi n + \theta)|}$. The collapse gives $\nu = 2$.

$k = F_{i_0}$ with $k\varphi - F_{i_0+1} = (-\varphi)^{-i_0}$ and the constraint that $\varphi^{-i_0} < \Delta/2 < \varphi^{-(i_0-1)}$. (If $k\varphi - F_{i_0+1} > 0$, we can simply take the next Fibonacci number with $k = F_{i_0+1}$.) The distance between defects sets the correlation length $\xi = F_i \approx \varphi^i \sim \Delta^{-1} \sim \delta W^{-1}$, which implies that $\nu = 1$. We checked numerically that $\nu = 1$ leads to a perfect finite size collapse of the probability for the RG to end with an insulating block (Fig. 3).

*Singular example:* The above argument for $\nu = 1$ suggests that more singular potentials can lead different exponents $\nu > 1$. When we wrote $\Delta \propto \delta W$, we implicitly assumed that the derivative of the potential is not singular at the point where $\{\varphi x + \theta/(2\pi)\} = \frac{\{\varphi\}}{2}$. However there exist some functions for which this is not true. For example, consider $\ell_n = f(\cos(2\pi\varphi n + \theta))$ with $f(x) = 2 + \text{sgn}\left(\cos(2\pi\frac{\{\varphi\}}{2}) - x\right)\sqrt{|\cos(2\pi\frac{\{\varphi\}}{2}) - x|}$. $f(x)$ is a bounded monotonic function in the range $(-1, 1)$ and hence flows to sequence under RG, but it has a singular derivative at $x = \cos 2\pi\frac{\{\varphi\}}{2}$. This implies that $\delta W \sim \sqrt{\Delta}$, so we expect $\nu = 2$ in this case, as confirmed by numerical simulations of the RG (Fig. 6).

## B. $\beta^I > 1$

As mentioned above, slightly increasing $\beta^I$ is an irrelevant perturbation to the sequence, and does not alter the transition values. But if $\beta^I$ is increased beyond a certain threshold, the asymmetric perturbation can dominate the sequence values and break down the sequence pattern (see Fig. 2). We get different self-similar sequences as $\beta^I$ is increased. The next self-similar sequence is for $3 < \beta^I < 12.75$, see Fig. 7. This sequence is self-similar under one Fibonacci step. More concretely if the sequence size is $F_{3(m+1)}$ then under RG, the sequence pattern will re-emerge when the system size becomes $F_{3m}$. To formalize this, we define the so-called substitution matrix of the sequence. If we define $w_1 = A^T$, $w_2 = A^I$, $w_3 = B^I$, $w_4 = B^T$, then the substitution matrix $M$ is given by $M_{ij} = \#w_i$ in inflation rule for $w_j$.

For the Fibonacci sequence discussed above for the case of $\beta^I$ close to 1, the inflation rules were $A^T \rightarrow A^T B^I A^T B^I A^T$, $B^I \rightarrow A^T B^I A^T$, and so on. This corresponds to the substitution matrix,

$$M = \begin{bmatrix} 3 & 0 & 0 & 2 \\ 0 & 3 & 2 & 0 \\ 2 & 0 & 0 & 1 \\ 0 & 2 & 1 & 0 \end{bmatrix}. \tag{9}$$

Let $w = [\#w_1, \#w_2, \#w_3, \#w_4]$ be the vector representing the number of various letters in the initial word sequence. Let $\lambda_i$ be the eigenvalues of $M$ (with $\lambda_1$ the largest value) and $v_i$ being the corresponding eigenvectors. We assume that we can write $w = \sum_i a_i v_i$ with some real coefficients $a_i$. The vector $Mw$ denotes the number of letters after a

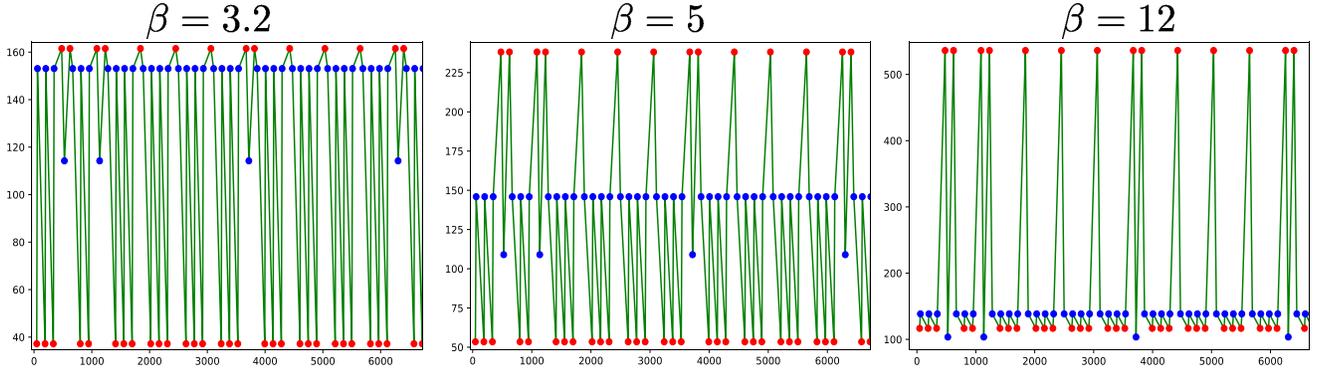

FIG. 7. Critical self-similar sequence after 3 Fibonacci step for different values of $\beta^I$ at the critical value of $\delta W \approx 1.97$. Red dot represent the thermal blocks and blue represents insulating. We see that on increasing $\beta^I$, the length of thermal blocks become greater and length of insulating blocks smaller (relative to one another). We checked that the sequence structure is preserved at all Fibonacci steps. Eventually on increasing $\beta^I$ further the sequence pattern will get destroyed, i.e the smaller thermal blocks will be larger than the largest insulating block.

single application of inflation rules. Thus after $n$ inflations, the total number of letters is given by,

$$
\begin{aligned}
N_n &= \sum_{i,j} (M^n)_{ij} w_j \\
&= \sum_{i,k} (\lambda_k^n a_k (v_k)_i \\
&= \lambda_1^n a_1 \sum v_1 \left( 1 + \left( \frac{\lambda_2}{\lambda_1} \right)^n \frac{a_2 \sum v_2}{a_1 \sum v_1} + ... \right) \\
&\approx \lambda_1^n a_1 \sum v_1,
\end{aligned}
$$

implying that the number of letters changes by a factor of $\lambda_1$ under single application of the inflation rule. For (9) $\lambda_1 = \varphi^3$, so that so that under one Fibonacci step, the number of blocks is decreased by a factor of $\varphi^3$.

Moving back to the sequence in Fig. 7 for $3 < \beta^I < 12.75$, the substitution matrix is given by

$$
M_2 = \begin{bmatrix} 2 & 0 & 0 & 1 \\ 0 & 4 & 3 & 0 \\ 1 & 0 & 0 & 0 \\ 0 & 3 & 2 & 0 \end{bmatrix}, \tag{10}
$$

with the largest eigenvalue given by $\varphi^3$. This shows that the sequence repeats itself after a single Fibonacci step. We remark that the range of $\beta^I$ over which the above sequence is well defined can be identified semi-analytically. We can obtain the lengths of the block after the 1st Fibonacci step as a function of $\beta^I$ by summing over the cosines (like we did for symmetric case). Then the lengths at further steps can be calculated as follows,

$$
\begin{bmatrix} A_{m+1}^T(\beta^I) \\ A_{m+1}^I(\beta^I) \\ B_{m+1}^T(\beta^I) \\ B_{m+1}^T(\beta^I) \end{bmatrix} = \begin{bmatrix} 2 & 0 & \beta^I & 0 \\ 0 & 4 & 0 & 3\beta_T \\ 0 & 3 & 0 & 2\beta_T \\ 1 & 0 & 0 & 0 \end{bmatrix}^m \begin{bmatrix} A_1^T(\beta^I) \\ A_1^I(\beta^I) \\ B_1^I(\beta^I) \\ B_1^T(\beta^I) \end{bmatrix}. \tag{11}
$$

Eq (11) can be calculated numerically and we can check that the sequence pattern, $A_m^T > A_m^I$, $B_m^T$, $A_m^I$ and $A_m^I > B_m^I$, is true at all Fibonacci steps for $3 < \beta^I < 12.75$.

As we increase $\beta^I$ further, we get different sequences with different periods of self-similarity. For $62 < \beta^I < 78$, we have another sequence whose substitution matrix is given by,

$$
M_3 = \begin{bmatrix} 2 & 3 & 3 & 1 \\ 0 & 4 & 3 & 0 \\ 1 & 17 & 14 & 0 \\ 0 & 16 & 13 & 0 \end{bmatrix}. \tag{12}
$$

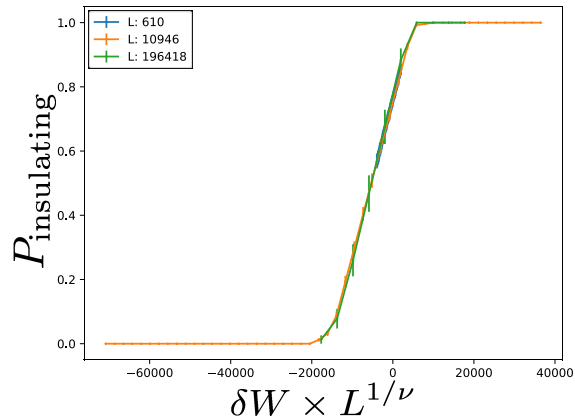

FIG. 8. Collapse for $\beta^I = 70$ with $\nu = 1$. For this value of $\beta^I$, the critical sequence repeats itself after two Fibonacci steps. Unlike Fig. 3, in which we had a crossing between systems with lengths of the form $F_{3m}$, in this case the crossing is seen only for system sizes $F_{3 \times 5} = 610$, $F_{3 \times 7} = 10946$ and $F_{3 \times 9} = 196418$, i.e for system sizes of the form $F_{3 \times (2m+1)}$.

The largest eigenvalue of $M_3$ is $\varphi^6$. This implies that the sequence repeats itself after two Fibonacci steps, i.e if the system size is $N = F_{3m}$ then under RG flow the sequence re-emerges when the system size is $F_{3(m-2)}$. This can be checked numerically (see Fig. 8). A nice finite size collapse for the transition is observed only for system sizes of the form $F_{3 \times (2m+1)}$, in contrast to the previous two transitions, for which collapses were obtained for system sizes of the form $F_{3m}$.

We have also observed more complicated sequences for larger values of $\beta^I$, with a period of 3 Fibonacci steps. Studying such sequences numerically is however very challenging as extremely large systems are needed. We expect that as one increases $\beta^I$, the transition is controlled by larger and larger sequences with complicated inflation rules and scaling of the form $\varphi^{3k}$. Our defect argument for $\nu = 1$ is however quite general, and applies generally to all these self-similar sequences.


\end{document}